\documentclass[preprint,
superscriptaddress,
amsmath,amssymb,aps,showkeys,showpacs,
twoside,final,secnumarabic,
nofootinbib]{revtex4-2}

\usepackage[paperwidth=205mm,paperheight=290mm,top=17mm,bottom=25mm,
inner=17mm,outer=17mm,
twoside]{geometry}

\usepackage{cmap} 
\usepackage[T1,T2A]{fontenc}
\usepackage[utf8]{inputenc}
\usepackage[russian,english]{babel}
\usepackage{color}
\usepackage{graphicx}
\usepackage{dcolumn}
\usepackage{bm} 
\usepackage[unicode=true,colorlinks=true,linkcolor=magenta, urlcolor=blue, citecolor = blue,breaklinks]{hyperref}
\usepackage{multirow}
\usepackage{url}
\usepackage{breakurl}
\DeclareGraphicsExtensions{.eps}

\newcount\issue
\newcount\Vol
\newcount\numb

\usepackage{fancyhdr} 
\def\Vol{\textbf{80}}
\def\numb{x}
\begin{document}

\title{ CONFERENCE SECTION \\[20pt]
Thermodynamics of charged black points in  vacuum nonlinear electrodynamics} 

\def\addressa{Physics Department, Moscow State University, Moscow, 119991 Russia}

\author{\firstname{V.A.}~\surname{Sokolov}}
\email[E-mail: ]{sokolov.sev@inbox.ru}
\affiliation{\addressa}

\received{xx.xx.2025}
\revised{xx.xx.2025}
\accepted{xx.xx.2025}

\begin{abstract}
The paper analyzes the thermodynamic properties of a special state of 
charged black holes, in which the event horizon becomes point-like 
and coincides with the singularity. It is shown that for such states, called black points, 
in contrast to Reissner-Nordstr\"{o}m black holes, 
the third law of thermodynamics in Planck's formulation is fulfilled. 
It is also shown that the formation of a black point is not a first-order phase transition. 
Considerable attention is paid to the special state of a black point 
with a doubly degenerate horizon, similar to the extreme state of a Reissner-Nordstr\"{o}m black hole.
\end{abstract}

\par
\keywords{vacuum nonlinear electrodynamics, charged black hole, extreme black point\\[5pt]}

\maketitle
\thispagestyle{fancy}

\section{Introduction}\label{intro}
Modification of the electromagnetic interaction sector by means of 
transition to nonlinear models of electrodynamics in vacuum opens 
up new possibilities for studying phenomena in extremely strong 
electromagnetic and gravitational fields. One of the reasons for such 
an extended use of nonlinear models is the possibility of naturally 
eliminating the problem of the divergence of the self-energy of the 
electrostatic field of point sources and solving the regularization
problem without resorting to renormalization schemes.
One of the first heuristic models to successfully solve the regularization problem,
Born–Ifeld electrodynamics~\cite{CH1_BornInfeld}, received support from superstring theory, 
where it was shown in~\cite{CH1_FradkinTseitlin} that this model corresponds to the action of 
an open string in the low-energy approximation for slowly varying fields.
A distinctive feature of the Born-Infeld model is the absence of the vacuum 
birefringence effect for electromagnetic waves in an external field, 
as well as the impossibility of forming shock wave fronts~\cite{CH1_BI_SWfree3}.

However, due to its heuristic nature, the Born-Infeld model is not the 
only possible nonlinear modification of electrodynamics in a vacuum.
The subsequent development of the idea of natural regularization of the 
field energy of point sources led to the emergence of a number of 
models with an analytically exact expression for the Lagrangian, some of 
which are presented in the Table~\ref{Tabe1}. 
As follows from the listed Lagrangians, when constructing the models, 
their authors, in addition to fulfilling the regularization condition, 
sought to observe the correspondence to Maxwell's electrodynamics with 
a particular choice of model parameters. However, this did not always
guarantee compliance with a number of fundamental principles, 
in connection with which the results obtained in \cite{CH2_EC_ShabadUsov} expressed in 
the form of restrictions on the Lagrangian of a general type, 
following from the principles of unitarity and causality, 
became an extremely powerful tool for selecting such models.

\begin{table}[htbp] \label{Tabe1}
\centering
\begin{tabular}{|c|c|}
\hline \rule[-1ex]{0pt}{4ex}
{\bf The model title} & {\bf The model Lagrangian}  \\ [2ex]
\hline \rule[-1ex]{0pt}{4ex}
{<<Modified>>  Born-Infeld model \cite{CH1_BI_Modif3}} & ${\cal L}=\beta^2\Big(1-\Big[1+\dfrac{2}{\beta^2}{\cal F}-\dfrac{1}{\beta^4}{\cal G}^2\Big]^p\Big)$ \\ [2ex]
\hline \rule[-1ex]{0pt}{5ex}
 Hofman-Infeld model \cite{CH1_HI_Model} & ${\cal L}=\dfrac{\beta^2}{4}[1-\eta({\cal F})-\ln\eta({\cal F})]$ \\ [2ex] 
\hline \rule[-1ex]{0pt}{5ex}
 <<Quadratic>> model \cite{CH1_POWLAW_Shabad} & ${\cal L}=-{\cal F}+\dfrac{\gamma}{2}{\cal F}^2$ \\ [2ex] 
\hline \rule[-1ex]{0pt}{5ex}
 <<Rational>> model \cite{CH1_Rational_Kruglov} & ${\cal L}=-\dfrac{{\cal F}}{1+2\beta {\cal F}}$ \\  [2ex]
\hline \rule[-1ex]{0pt}{5ex}
{<<Double parametric rational>> model} \cite{CH1_DParam_Kruglov} & ${\cal L}=-{\cal F}-\dfrac{{a\cal F}}{1+2\beta {\cal F}}$ \\ [2ex]
\hline \rule[-1ex]{0pt}{5ex}
 <<Logarithmic>> model \cite{CH1_LogarithmicModel_Gaete} & ${\cal L}=-\beta^2\ln
\Big(1+\dfrac{{\cal F}}{\beta^2}-\dfrac{{\cal G}^2}{2\beta^4}\Big)$ \\  [2ex]
\hline \rule[-1ex]{0pt}{4ex}
{<<Double logarithmic>>  model} \cite{CH1_DLogarithm_Gulu} & 
{${\cal L}=\dfrac{1}{2\beta}\Big[(1-{\cal Y}\ln(1-{\cal Y})+(1+{\cal Y}\ln(1+{\cal Y}) \Big]$} \\ [2ex]
\hline \rule[-1ex]{0pt}{5ex}
 <<Exponential>> model \cite{CH1_BI_Modif3} & ${\cal L}=\beta^2(e^{-X/\beta^2}-1)$  \\ [2ex]
\hline \rule[-1ex]{0pt}{5ex}
 <<Arctan>> model \cite{CH1_ATANMOD_Kruglov}  & 
${\cal L}=-\dfrac{1}{\beta}\arctan(\beta {\cal F})$  \\ [2ex]
\hline \rule[-1ex]{0pt}{5ex}
 <<Arcsine>> model \cite{CH1_ASINMOD_Kruglov} & ${\cal L}=-{\cal F}+
\dfrac{C}{\beta}\arcsin(\beta{\cal F})+\dfrac{\gamma}{2}{\cal G}^2$  \\ [2ex]
\hline 
\end{tabular} 
\caption{Some models of nonlinear vacuum electrodynamics,
possessing the property of regularization of the energy of the electrostatic field of a point charge in pseudo-Euclidean space-time. In all expressions, the original designations of the authors are preserved: ${\cal F}=1/4 F_{ik}F^{ik}$ -- is a scalar and  ${\cal G}=1/4 F_{ik}{}^{\ast}F^{ik}$ -- is a pseudoscalar  of the electromagnetic field tensor;
$\beta$,$\gamma$, $a$, $p$, $C$ -- are model parameters, and also used the notations 
$X={\cal F}-\dfrac{{\cal G}^2}{2\beta^2}, \quad {\cal Y}=\sqrt{\sigma\beta^2 {\cal G}^2-2\beta{\cal F}}$.}
\end{table}

For most of these models, the predictive base is identical, differing only in details in the quantitative description of effects, however, in some cases, extremely non-trivial properties of the theory are possible.
For example, in the "quadratic" model of electrodynamics \cite{CH1_POWLAW_Shabad}, regularization is manifested only partially: the electrostatic field strength
at the center of a point charge is infinite, while the field energy is limited.

In the electrovacuum solutions for the field of point sources (electric or dion charges)
in the General Theory of Relativity (GR), the regularizing properties of the listed models do not manifest themselves at all \cite{CH1_BI_Type_BH_Kruglov, CH1_EXP_Type_BH_Kruglov, CH1_Rational_Type_BH_Kruglov, CH1_ATAN_Type_BH_Kruglov, CH1_DLOGARITHM_Type_BH_Gulu}.
Nevertheless, some of them lead to extremely unusual singular solutions. For example, for the model of <<logarithmic>> electrodynamics minimally related to GR, in \cite{CH1_BP_Soleng} a solution was obtained for the field of a point charge called the <<black point>>. For this solution, for a certain set of parameters, the event horizon degenerates into a point coinciding with the position of the singularity. In the following, we will show a similar possibility
for Born-Infeld electrodynamics.

The solution to the regularization problem for point massive self-gravitating sources, as well as in flat space-time, can also be found in terms of nonlinear vacuum electrodynamics models.
One of the first successful implementations of this approach 
was obtained by J. Barden \cite{CH1_REGBH_Bardeen},
and led to the emergence of a new class of solutions 
to the equations of general relativity, called regular black holes.

The space-time of these objects has coordinate singularities corresponding 
to horizons, but lacks true singularities of the curvature tensor invariants.
Table~\ref{Tabe2} lists the Lagrangians of the most 
well-known models of nonlinear vacuum electrodynamics that admit 
solutions in the form of regular black holes in GR.

\begin{table}[htb]
\centering
\begin{tabular}{|c|c|}
\hline \rule[-1ex]{0pt}{4ex}
{\bf The model title} & {\bf The Lagrangian}  \\ [2ex]
\hline \rule[-1ex]{0pt}{6ex}
 Bardeen model \cite{CH1_REGBH_Bardeen} & ${\cal L}=\dfrac{4p}{a}\dfrac{ (aJ_2)^{5/4}}{[1+\sqrt{aJ_2}]^{1+p/2}}$ \\ [2ex]
\hline \rule[-1ex]{0pt}{6ex}
Hayword model \cite{CH1_REGBH_Hayword}& ${\cal L}=\dfrac{4p}{a}\dfrac{ (aJ_2)^{\frac{p+3}{4}}}{[1+(aJ_2)^{p/4}]^2}$ \\ [2ex]
\hline \rule[-1ex]{0pt}{6ex}
Bronnikov model \cite{CH1_REGBH_Bronnikov_Model}& ${\cal L}=-\dfrac{J_2}{\cosh^2(a|J_2/2|^{1/4})}$ \\ [2ex]
\hline \rule[-1ex]{0pt}{6ex}
 Ay\'{o}n--Beato--Garc\'{i}a model \cite{CH1_REGBH_Ayon} & 
$ \begin{array}{l}
 {\cal L}=P\dfrac{(1-8\sqrt{-2q^2P}-6q^2P)}{(1+\sqrt{-2q^2P)^4}}- \\ [2ex]
-\dfrac{3}{4q^2s}\dfrac{(-2q^2P)^{5/4}(3-2\sqrt{-2q^2P})}{(1+\sqrt{-2q^2P})^{7/2}} 
 \end{array} $
 \\  [3ex]
\hline \rule[-1ex]{0pt}{5ex}
Dymnikova model \cite{CH1_REGBH_Dymnikova_Model}& ${\cal L}=\dfrac{P(1-\alpha\sqrt{-P})}{(1+\alpha\sqrt{-P})^3}$ \\ [2	ex]
\hline 
\end{tabular} 
\caption{Models of electrodynamics that admit electrovacuum solutions in the form of regular black holes; Constant parameters $a$, $p$, $q$, $s$. For the last two models, the Lagrangian is not explicitly defined, since the invariant
$P=-({\partial {\cal L}}/{\partial J_2})^2 J_2$ is expressed through $J_2=F_{ik}F^{ki}$ implicitly.}\label{Tabe2}
\end{table}
A special place is occupied by the model of A\'{y}on-Beato and Garc\'{i}a \cite{CH1_REGBH_Ayon}, the authors of which were the first to obtain a solution for a regular black hole, in the limit of a weak field, which passes to the corresponding solution in Maxwell's electrodynamics.
In this paper we focus on the properties of charged 
black holes in the black point state.

Considerable emphasis will be placed on the study of the Born-Infeld model 
as the classical and most justified. The other models listed above have many similar properties, 
differing just in quantitative values. In Sec.\ref{Sec:Genaral_NED_Thermodynamics}, 
the thermodynamic features of the black points revealed for the Born-Infeld model will be translated for the general case of Lorentz-invariant nonlinear vacuum electrodynamics.

\section{Einstein-Born-Infeld black hole}
The variation of the action functional in the Einstein-Born-Infeld model leads to the 
equations of the electromagnetic and gravitational fields of the following form:
\begin{gather}
\nabla_j F_{nm}+\nabla_n F_{mj}+\nabla_m F_{jn}=0, \label{BI_Equat_QM}\\ \quad 
\nabla_n\Bigg\{{\big(2-a^2J_2\big)F^{kn}+2a^2F^{kn}_{(3)}\over
\sqrt{4-2a^2J_2-a^4J_4+a^4J_2^2/2\ \ } }\Bigg\}=-{4\pi\over c}j^k, \nonumber \\
R_{ik}-{1\over 2}g_{ik}R = \label{Grav_BI_Equat_QM} \\ 
=\frac{2\Big\{F^{(2)}_{ik}-g_{ik}\Big[
a^2J_{2}+2\sqrt{1-a^2J_2/2-a^4J_4/4+a^4J_2^2/8}-2\Big]/2a^2\Big\}}{\sqrt{1-a^2J_2/2-a^4J_4/4+a^4J_2^2/8}}, \nonumber
\end{gather}
where $F^{(2)}_{ik}=F_{im}F^{m\ \cdot}_{\cdot\ \ k}$ and
$F_{(3)}^{kn}=F^{kp}F_{pm}F^{mn}$  are the second and third power of the 
electromagnetic field tensor, $J_2=F_{ik}F^{ki}$ and 
$J_4=F_{ik}F^{kl}F_{lm}F^{mi}$ are the electromagnetic field invariants, 
and $a$ is Born-Infeld model parameter. 
The solution to the problem of the field of a self-gravitating static point 
source with mass $M$ and electric charge $Q$ leads to the metric function 
and the electric field strength of a charged black hole in the following form:
\begin{equation}
F_{01}=E={Q\over \sqrt{r^4+a^2Q^2}}, \qquad 
g_{00}=1-{2M\over r}+{2\over a^2 r}\int\limits^\infty_r
\big[\sqrt{\eta^4+a^2Q^2}-\eta^2\big]d\eta, \label{SolFinalQM}
\end{equation}
which turns into the well-known Reissner-Nordstr\"{o}m solution in Maxwell electrodynamics 
in the limit $a\to 0$.
Some properties of the metric \eqref{SolFinalQM} 
were investigated in \cite{S2011, S2016c}, in particular it was established:
the minimum of the metric function $g_{00}(r_m)=-b^2$ exists if the condition is satisfied
$Q^2>{(1+b^2)a^2/4}$; if the constant $b=0$, then the equation $g_{00}(r_s)=0$ has
only one double root, coinciding
with the minimum point; the metric function $g_{00}$ monotonically 
increases when the condition is satisfied:
$M>{(\pi|Q|)^{3/2} / 3\sqrt{a}\Gamma^2(3/4)}$.

The listed features allowed us to establish two important properties of space-time:
the number of horizons depends on the parameters of the black hole and with some 
combination of them it is possible to eliminate the inner horizon and 
as well as the possibility of a black hole transitioning to a black point state 
when the  horizon tends to singularity position.

\begin{figure*}[tbp]
\includegraphics[width=.49\textwidth,clip]{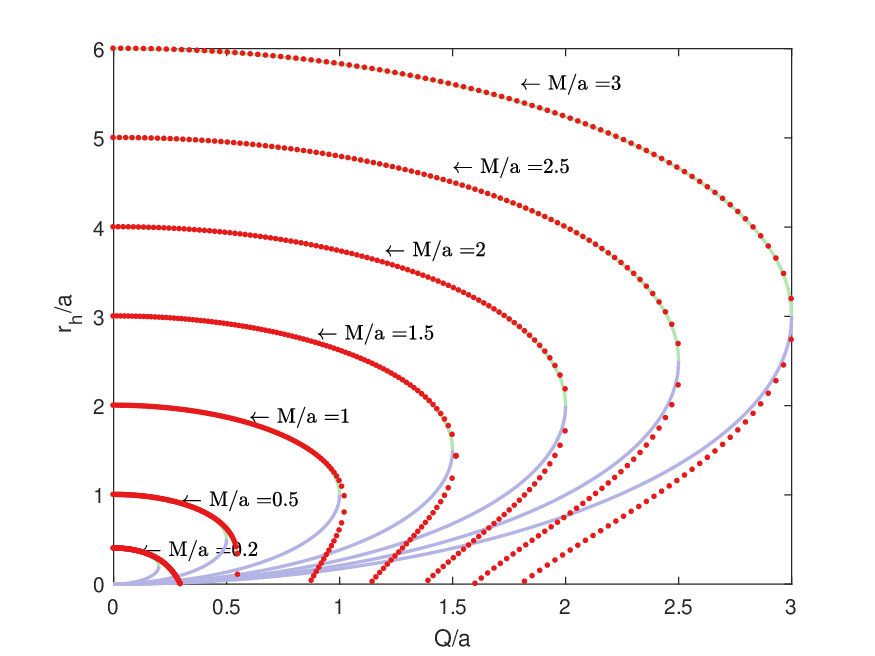}
\hfill
\includegraphics[width=.49\textwidth]{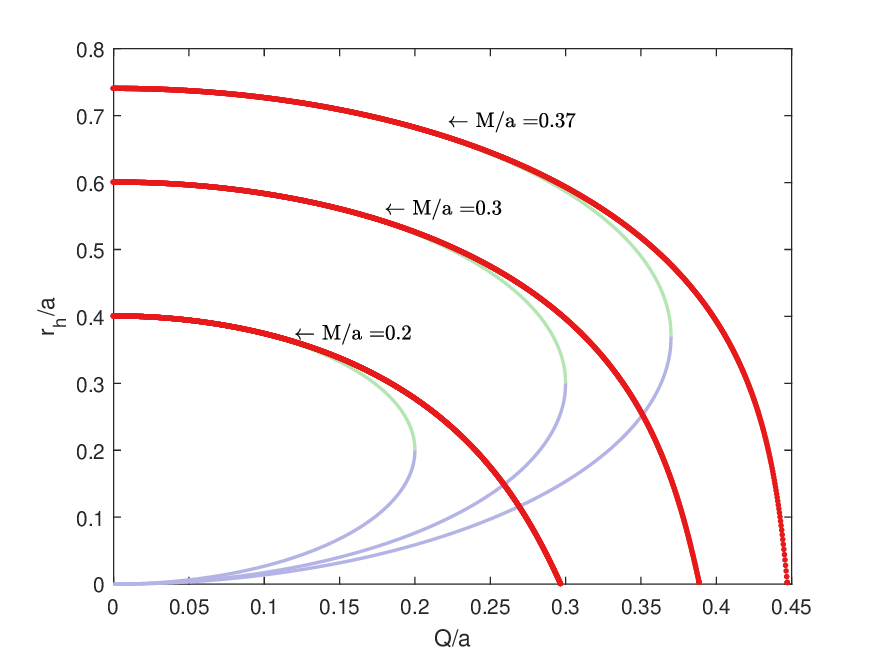}
\caption{ Dependence of the horizon radius on the electric charge of a 
black hole for different values of its mass.
The solid green and blue lines on the graph show the radii 
of the horizons for Reissner-Nordstr\"{o}m metric.
The left panel shows the horizon radius as the charge changes over a wide range. 
On the right panel is shown the radius of the horizon for cases of
formation of a black dot state.
\label{HorizonRadius_wideQ}}
\end{figure*}
It is noteworthy that the existence of a state with a doubly 
degenerate horizon in contrast to the extremal state of a
Reissner-Nordstr\"{o}m black hole is not 
possible for all values of the mass.
For the existence of the extreme state of the Einstein-Born-Infeld black hole, 
its mass and charge $Q'$ must satisfy the equation:
\begin{equation}
1-\frac{2M}{\sqrt{Q'^2-a^2/4}}+
\frac{2Q'^2}{\sqrt{Q'^2-a^2/4}}\int\limits_0^{{1}/{\sqrt{Q'^2-a^2/4}}}
\frac{d \xi}{1+\sqrt{1+a^2Q'^2\xi^4}}=0, \label{Extreme_state_equation}
\end{equation}
the solution of which becomes possible only if:
\begin{equation*}
M\geq {M_{cr}}=\frac{a^2}{4}\int
\limits_0^\infty {d \eta \over 1+\sqrt{1+a^4\eta^4/4}}=
\frac{1}{3\Gamma^2(3/4)}\Big(\frac{\pi}{2}\Big)^{3/2}a\simeq 0.437 a, \quad Q'\geq Q_{cr}=a/2.
\end{equation*}
In the case $M<M_{cr}$ there can be only one horizon, and a black point can be 
formed when the horizon radius becomes zero for the corresponding charge value. 
This case is realized in the right panel of Figure~\ref{HorizonRadius_wideQ}.
For $M>M_{cr}$, the existence of two horizons (for some values of charge) 
becomes possible, as well as the emergence of an extreme state of the black 
hole, in which these horizons coincide. The charge $Q'$ necessary for
the formation of such a state for a given mass can be obtained from the solution
of the equation~\eqref{Extreme_state_equation}.
Finally, the case $M=M_{cr}$ and $Q=a/2$ corresponds to an extreme black point,
a state in which the horizon $r_h=0$ is a second-order zero of the metric
function and coincides with the singularity.
In what follows, we will focus on examining the
thermodynamic properties of the black point states when $M\leq M_{cr}$.

For comparison, we will describe some thermodynamic
features of black holes of different types.
For a Schwarzschild black hole with mass $M\to 0$, 
the temperature $T={1}/{8\pi M}$ increases indefinitely, and
the entropy $S=4\pi M^2$ and the heat capacity $C=-16\pi M^2$ tend to zero.
The temperature divergence is usually explained by the intense evaporation of low-mass black holes.

The analogous thermodynamic parameters for a Reissner-Nordstr\"{o}m 
black hole with a horizon $r_h=M+\sqrt{M^2-Q^2}$, are given by the expressions:		
\begin{equation}
T=\frac{1}{4\pi r_h}\Big(1-\frac{Q^2}{r_h^2}\Big), \qquad S=\pi r_h^2, \qquad
C=2\pi r_h^2\Big(\frac{r_h^2-Q^2}{3Q^2-r_h^2}\Big).
\end{equation}
In the extreme state at $Q=M$, the horizon radius and entropy $S\to \pi M^2$ remain finite,
while the temperature tends to zero, which leads to a violation of the 
third law of thermodynamics in Planck's formulation.
Also, the heat capacity experiences a break at a charge value
of $Q^2=3/4M^2$, which corresponds to a first-order phase transition.
It seems extremely interesting to find out which of the noted
properties of the Schwanzschild and Reissner-Nordstr\"{o}m black
holes will be inherited by the Einstein-Born-Infeld black 
hole in the state with one horizon, as well as in the state
of an ordinary and extreme black point.

\begin{figure*}[tbp]
\includegraphics[width=.49\textwidth,clip]{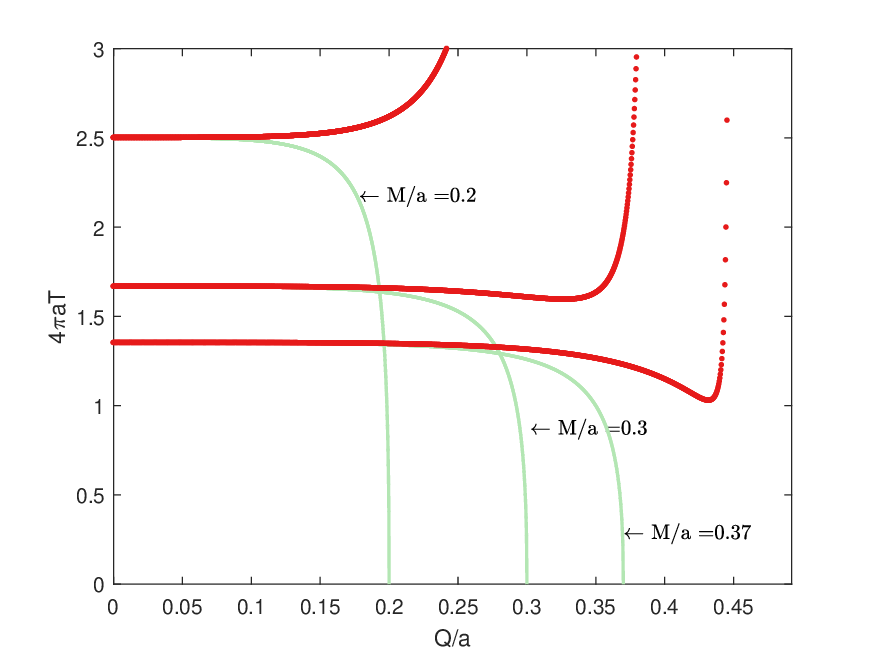}
\hfill
\includegraphics[width=.49\textwidth]{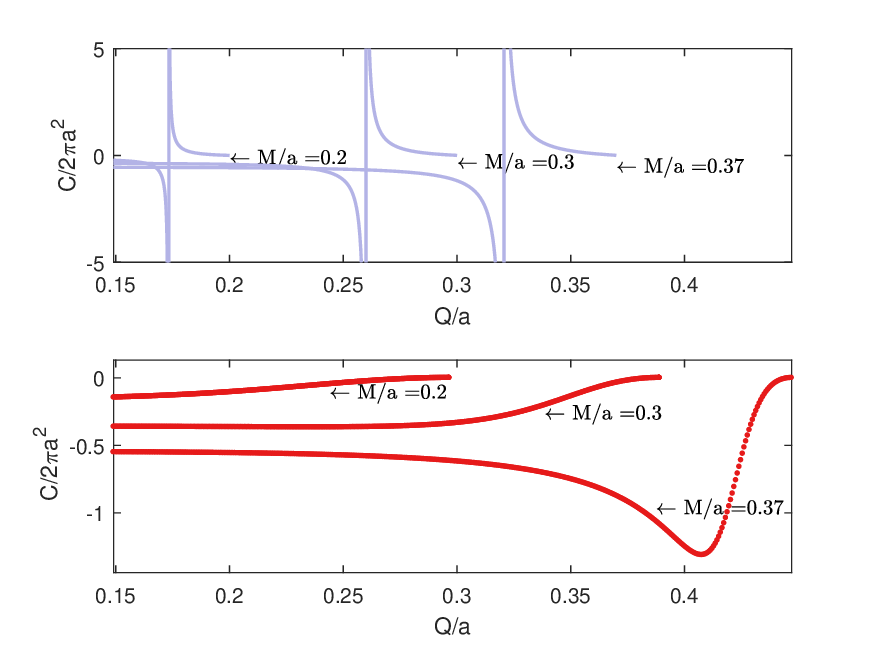}
\caption{Dependence of the normalized temperature $T$ and heat 
capacity $C$ on the charge of the black hole at $M<M_{cr}$. 
The data are plotted as solid green and blue lines for the 
Reissner-Nordstr\"{o}m black hole, and the corresponding 
quantities for the Einstein-Born-Infeld
black hole are plotted in red.
}\label{TempHeatCapacity}
\end{figure*}

Using expressions \eqref{SolFinalQM}, one can obtain in explicit form the 
dependence of temperature, entropy and heat capacity on the radius
of the horizon; the horizon itself can be determined numerically 
for a given value of the mass of the black hole:

\begin{equation}
T=\frac{1}{4\pi}\frac{\partial_r g_{00}}{|g_{00}g_{11}|}\Big|_{r_h}=\frac{1}{4\pi r_h}\Big(1-\frac{2Q^2}{r_h^2+\sqrt{r_h^4+a^2Q^2}}\Big), \qquad S=\pi r_h^2, \label{Temp_BI}
\end{equation}
\begin{equation}
C=T\Big(\frac{\partial S}{\partial T}\Big)=2\pi r_h^2
\frac{\sqrt{r_h^4+a^2Q^2}(r_h^2-Q^2)+a^2/4}{\sqrt{r_h^4+a^2Q^2}(Q^2-a^2/4)+r_h^2(2Q^2-r_h^2)}. \label{HeatCapacity_BI}
\end{equation}
Figure~\ref{TempHeatCapacity} shows the dependence of temperature and heat capacity on the black hole charge, for mass values less than the critical mass required to form the inner horizon. The left panel shows the dependence of temperature on the normalized charge. Solid green lines correspond to the Reissner-Nordstr\"{o}m black hole, the corresponding dependence for the Einstein-Born-Infeld black hole is plotted in red.

As the charge approaches the maximum allowable value for a 
black hole of a given mass, the temperature behaviour for these
two types of black holes differs qualitatively:
for a Reissner-Nordstr\"{o}m black hole, the temperature tends to zero,
whereas for an Einstein-Born-Infeld black hole, as it approaches
the state of an ordinary black point, the temperature increases
indefinitely, as for a Schwarzschild black hole.

The right panel shows the dependence of the heat capacity on the 
charge for the Reissner-Nordstrom black hole (at the top) 
and for the Einstein-Born-Infeld black hole (at the bottom).
As can be seen from the simulation results, the heat capacity 
of the Einstein-Born-Infeld black hole at $M<M_{cr}$ is continuous and 
tends to zero in the black point state, so the evolution to this 
state itself is not associated with a first-order phase transition.

The case of an extreme black point requires special consideration.
To do this, we perform the expansion of expressions \eqref{Temp_BI}, \eqref{HeatCapacity_BI} 
near the state with zero horizon radius $r_h$:
\begin{equation}
T\Big|_{r_h\to 0}\simeq \frac{1}{4\pi a}
\Big(\frac{{a-2|Q|}}{r_h}+\frac{2r_h}{a}-\frac{r_h^3}{a^2|Q|}+\frac{r_h^7}{4a^4|Q|^3}+{\cal O}(r_h^{11})\Big),
\label{T_BlackPoint_BI}
\end{equation}
\begin{equation}
C\Big|_{r_h\to 0}\simeq -2\pi r_h^2
\Big(1+\frac{4r_h^2}{a{(a-2|Q|)}}-
\frac{4(a-4|Q|)r_h^4}{a^2|Q|{(a-2|Q|)^2}}+{\cal O}(r_h^{8})\Big).
\label{C_BlackPoint_BI}
\end{equation}

The appearance of a term with a negative power of the 
horizon radius in expression \eqref{T_BlackPoint_BI} leads to a divergence 
of temperature when approaching the ordinary black point state, 
which was noted earlier in the numerical simulation. 
At the same time, the heat capacity is devoid of such 
a feature and is equal to zero in this state.

In the case of an extreme black point $M=M_{cr}$  and $|Q|=a/2$, 
the properties change: the term with the negative power of the horizon 
radius disappears, and the temperature becomes regular and tends to zero. 
An uncertainty arises in the expression for heat capacity, which nevertheless 
leads it to a zero value.

It is noteworthy that in the extreme black point state both temperature
and entropy $S=\pi r_h^2 \to 0$ are also tend to zero, which corresponds
to the third law of thermodynamics formulated by Planck.
In the next section, the thermodynamic properties obtained for 
the black points in Einstein-Born-Infeld model will be generalized
to the case of an arbitrary vacuum nonlinear electrodynamics.	

\section{Thermodynamics of black points in generalized models}\label{Sec:Genaral_NED_Thermodynamics}
Let us consider general form of nonlinear vacuum electrodynamics with the Lagrangian ${\cal L}$,
minimally coupled with Einstein's general theory of relativity, 
for which the action functional is:
\begin{equation}
S=\int\sqrt{-g}d^4x\Big[-\frac{R}{16\pi}+{\cal L}(J_2, J_4)-A_k j^k\Big], \label{Act_GeneralForm}
\end{equation}
where $J_2=F_{ik}F^{ki}$ and $J_{4}=F_{ik}F^{kl}F_{lm}F^{mi}$ are 
the invariants of the electromagnetic field tensor, $R$ is the Ricci scalar, $g$ 
is the determinant of the metric, $A^k$ and $j^k$ are the 4-vectors of the potential and current density, respectively. 

As usual, by varying the action functional over the metric, we obtain the Einstein equations $R_{ik}-{Rg_{ik}}/{2}=8\pi T_{ik}$ with the energy-momentum tensor for nonlinear vacuum electrodynamics in the following form:
\begin{eqnarray}\label{Str_Energ_Gen}
T_{ik}=\frac{2}{\sqrt{-g}}\frac{\delta (\sqrt{-g}{\cal L})}{\delta g^{ik}}&=&
4\Big[\frac{\partial {\cal L}}{\partial J_2}+
J_2\frac{\partial {\cal L}}{\partial J_4}\Big]F_{ik}^{(2)}+
\Big[(2J_4-J_2^2)\frac{\partial {\cal L}}{\partial J_4}-{\cal L}\Big]g_{ik},
\end{eqnarray}
where $F_{ik}^{(2)}=F_{im}F^{ml}g_{lk}$ is the second power of the electromagnetic field tensor.
Variation over the variables of the electromagnetic field leads to 
the equations of nonlinear vacuum electrodynamics:
\begin{eqnarray}
\frac{1}{\sqrt{-g}}\frac{\partial \sqrt{-g} Q^{kn}}{\partial x^n}=-{j^k},  \qquad 
Q^{kn}=4\frac{\partial {\cal L}}{\partial J_2}F^{kn}
+8\frac{\partial {\cal L}}{\partial J_4} F_{(3)}^{kn},  \label{Q_tens}
\label{Em_Field_eq} \
\end{eqnarray}
in many ways similar to the equations of the electromagnetic field in an effective medium,
the properties of which are described by the auxiliary tensor $Q^{kn}$
which depends on the components of the tensor $F^{ik}$, 
its third power $F_{(3)}^{kn}=F^{kl}F_{lm}F^{mn}$ and the 
choice of the model of nonlinear vacuum electrodynamics.

We will calculate the field of a stationary self-gravitating
point source, with mass $M$ and electric charge $Q$, the space-time line element  for which and the current density can be written as:
\begin{equation}\label{Line_elem}
ds^2=e^{2\alpha(r)}dt^2-e^{-2\alpha(r)}dr^2-r^2(d\theta^2+\sin^2\theta d\varphi^2), 
\qquad j^{k}=\frac{Q}{4\pi r^2}\delta_0^k\delta(r),
\end{equation} 
where  $\delta_i^k$ is Kronecker and $\delta(r)$ is Dirac deltas. 
Let us write down the components of the electromagnetic field tensor for 
electric field with the strength $E$ of such a source and,
using the metric given by relations \eqref{Line_elem},
we calculate its powers and the invariants:
\begin{align}\label{F_tens}
&F_{ik}=E(r)\{\delta_i^0\delta_k^1-\delta_i^1\delta_k^0\},
\qquad 
F_{(2) i\cdot}^{\cdot n}=F_{ik}F^{kn}=E^2\{\delta_i^0\delta_0^n+\delta_i^1\delta_1^n\}, 
\\ 
&F_{(3)}^{mn}=F^{mi}F_{(2) i\cdot}^{\cdot n}=
-E^3\{\delta_0^m\delta_1^n-\delta_1^m\delta_0^n\}, \qquad J_2=2E^2, \qquad  J_4=J_2^2/2. \nonumber
\end{align}
Using expressions \eqref{Line_elem}, \eqref{F_tens} it is easy to obtain a general solution to equations
\eqref{Em_Field_eq}, describing the electric field of a point charge in an arbitrary model
of nonlinear vacuum electrodynamics:
\begin{equation}
4\Big[\frac{\partial {\cal L}}{\partial J_2}+2E^2\frac{\partial {\cal L}}{\partial J_4}\Big]E\Big|_{J_4=J_2^2/2}
=\frac{Q}{4\pi r^2}. \label{Electric_field_Solution}
\end{equation}

Einstein's equations for the spherically symmetric vacuum field under
consideration are reduced to only one non-trivial 
linearly independent equation:
\begin{eqnarray}\label{Einstein_eq_final_form}
e^{2\alpha}\Big[\frac{2\alpha'}{r}+\frac{1}{r^2}\Big]-\frac{1}{r^2}=-8\pi T_{0 \cdot}^{\cdot 0}, 
\end{eqnarray}  
the solution of which can be easily obtained in quadratures using the standard substitution:
\begin{equation}
g_{00}=e^{2\alpha}=1-\frac{2M}{r}+\frac{\Phi(r)}{r}, 
\qquad \Phi(r)=8\pi\int\limits_r^\infty T_{0\cdot}^{\cdot 0} r^2 dr, \label{Metric_Function_Solution}
\end{equation}	
where the component of the stress-energy tensor \eqref{Str_Energ_Gen} can be simplified 
using \eqref{Electric_field_Solution} 
and explicitly expressed through the Lagrangian of the nonlinear vacuum electrodynamics
and the electric field strength:
\begin{equation}\label{StessEnergy_Component}
T_{0\cdot}^{\cdot 0}=4\Big[\frac{\partial {\cal L}}{\partial J_2}+
2E^2\frac{\partial {\cal L}}{\partial J_4}\Big]E^2\Big|_{J_4=J_2^2/2}-{\cal L}=\frac{QE(r)}{4\pi r^2}-{\cal L}.
\end{equation}

To calculate the thermodynamic parameters, 
we determine the horizon radius from the solution of the equation
$g_{00}(r_h)=0$, the consequence of which and \eqref{Metric_Function_Solution} 
is the expression $\Phi(r_h, Q)=2M-r_h$.
In the case of a doubly degenerate horizon,
an additional requirement must also be met $\partial_r g_{00}(r_h)=0$, which leads to 
$1+\Phi'(r_h, Q)=0$, where the prime denotes the derivative over the horizon radius. 

In the case of a black point, in all the obtained relations
one should take the limit $r_h\to 0$, therefore, 
the expansion of the horizon equation in a series leads to:
\begin{equation}
\Big[\Phi(0, Q)-2M\Big]+\Big[1+\Phi'(0, Q)\Big]r_h+\frac{1}{2}\Phi''(0, Q)r_h^2+\cdots =0. \label{HorizonEquationDec}
\end{equation}
For an ordinary black point, it is sufficient for only the expression in the 
first square brackets to become zero, which leads to an integral equation that allows,
for a given mass value, to determine the charge corresponding to the state of the black point:
\begin{equation}
M=4\pi\int\limits_0^\infty T_{0\cdot}^{\cdot 0}(Q,r) \: r^2 dr. 
\end{equation}
It is noteworthy that in this state the energy of the electromagnetic field 
completely corresponds to the mass, which will obviously also be true for the extreme black point.

For an extreme black point, with a doubly degenerate horizon, both square brackets must already vanish. 
The equality to zero of the second square bracket leads to an equation for determining the charge $Q_{cr}$,
the substitution of which into the expression following from the equality to zero of the first bracket 
leads to the value of the critical mass $M_{cr}$ 
for which the existence of an extreme black point becomes possible:
\begin{equation}
1+\Phi'(0, Q_{cr})=1-8\pi T_{0\cdot}^{\cdot 0} r_h^2\Big|_{r_h \to 0}=0 , \qquad  2M_{cr}=\Phi(0, Q_{cr}). \label{ExtremeBlackPoint_GeneralCase}
\end{equation}
Based on expression \eqref{HorizonEquationDec} 
it is possible to assume the existence of black points with a higher degree of degeneration,
for which, in addition to relations \eqref{ExtremeBlackPoint_GeneralCase}, 
the higher derivatives of the function $\Phi$ should also vanish at $r_h \to 0$. 
In this case, equations \eqref{ExtremeBlackPoint_GeneralCase}, as before, 
will allow one to determine the critical charge and mass of such a black point, 
and the vanishing of the corresponding derivatives of the function $\Phi$ 
will lead to restrictions for the Lagrangian of the model. 
Let us move on to considering thermodynamic properties.

Using the solution of the metric function \eqref{Metric_Function_Solution}, we determine the 
temperature and heat capacity of a black hole in the general case, 
similar to how this was done earlier in \eqref{Temp_BI} and \eqref{HeatCapacity_BI} 
for the Einstein-Born-Infeld black hole:
\begin{equation}
T=\frac{1}{4\pi r_h}\Big[1+\Phi'\Big]=
\frac{1}{4\pi r_h}\Big[1-8\pi r_h^2 T_{0 \cdot}^{\cdot 0}\Big], \label{T_General_NED}
\end{equation}
\begin{equation}
C=-2\pi r_h^2\Big(\frac{1+\Phi'}{1+\Phi'-r_h \Phi''}\Big)=
-2\pi r_h^2\Big(\frac{1-8\pi r_h^2 T_{0\cdot}^{\cdot 0}}{1+8\pi r_h^2(r_h T_{0\cdot}^{\cdot 0})'}\Big). 
\label{C_General_NED}
\end{equation}
As we approach the state of an ordinary black point $r_h\to 0$, 
the temperature increases  to infinity $T\sim {\cal O}(1/r_h)$, 
and the heat capacity tends to zero $C\sim -2\pi r_h^2$.

In the case of the extreme black point $1+\Phi'\to 0$, an uncertainty 
arises in the expression for the temperature \eqref{T_General_NED}, which, with the additional
requirement $\Phi''\to 0$, leads to a result similar to the 
Einstein-Born-Infeld model -- the tendency of the temperature to zero.
However, when this condition is met, the heat capacity also 
acquires uncertainty in the state of the extreme black point, 
for the resolution of which it is sufficient to require
that $r_h\Phi'' \sim \bar{\bar{o}} \big(\Phi'''\big)$.	
For the Born-Infeld model
$\Phi''' \to 4/a^2$ at $r_h \to 0$ and 
this condition will be satisfied.	
If we assume that the Lagrangian and the electric field strength are 
an even functions of $r_h$, then, 
based on the expression~\eqref{StessEnergy_Component}, we can expect
similar properties for other
models of nonlinear vacuum electrodynamics.

\section{Conclusion}
The article examined the features of the space-time and the thermodynamic 
properties of a special state of charged black holes in the form of black points.
Using the Einstein-Born-Infeld model as an example, the dependences of 
the temperature and heat capacity of black points on the charge were studied 
for different mass values. 
It was found that when approaching the state of an ordinary black point, 
the temperature increases infinitely, and the heat capacity is continuous 
and tends to zero. Such behaviour of these parameters turns out to be closer
to the properties of a Schwarzschild black hole neater than
a charged Reissner-Nordstr\"{o}m black hole.

For the extreme black point state, the regular behaviour of temperature and 
heat capacity is shown, and both of these parameters, as well as entropy, tend
to zero when approaching this state, which indicates the fulfillment of the 
third law of thermodynamics in Planck's formulation, 
in contrast to the extreme Reissner-Nordstr\"{o}m black hole.
The obtained properties are generalized to the case of an arbitrary 
Lorentz-invariant model of nonlinear vacuum electrodynamics. 
The conditions for the existence of ordinary and extreme black 
points are established~\eqref{HorizonEquationDec}, 
for which the temperature and heat capacity are also analyzed.

According to~\eqref{T_General_NED} and~\eqref{C_General_NED}, 
for ordinary black points in general nonlinear vacuum electrodynamics models, 
the temperature diverges while the heat capacity tends to zero.
For the extreme black point, the temperature, like the heat capacity, tends to zero, 
provided that the second derivative with respect to the horizon radius of the function $\Phi$,
which describes the metric in general solution \eqref{Metric_Function_Solution}, 
tends to zero in this state, and the condition for 
the third derivative of this function, noted in the Section \ref{Sec:Genaral_NED_Thermodynamics}
is also satisfied. The fulfillment of these conditions is easy to verify 
for particular models and seems quite realistic in the general case.
Thus, for a black hole in the state of an extreme black point, the third 
law of thermodynamics is fulfilled, and the regularizing properties of
nonlinear electrodynamics are fully manifested, which makes this state 
extremely interesting and promising for further research.

\begin{acknowledgments}
I would like to express my deep gratitude to my colleagues: 
K.A. Sveshnikov, D.A. Slavnov, A.E. Lobanov, Yu.M. Loskutov, A.G. Loskutova,
whose memories will always be inspiring. The study was conducted
under the state assignment of Lomonosov Moscow State
University.
\end{acknowledgments}

\bibliography{BP_Bibliography.bib}

\end{document}